
\vsize=7.5in
\hsize=6.6in
\hfuzz=20pt
\tolerance 10000

\baselineskip 12pt plus 1pt minus 1pt
\pageno=0

\def\){]}
\def\({[}

\def\rjustline#1{\line{\hss#1}}

\rjustline{WIS-93/8/Jan-PH}

{\twelvepoint\bf
\centerline{Interference Effects in B-Decays to Flavor-Mixed Neutral
Mesons}
\centerline{Clues to Small Amplitudes and CP-Violation}}
\vskip  .40in
\twelvepoint
\centerline{Harry.~J.~Lipkin}
\smallskip
\centerline{Department of Nuclear Physics, Weizmann Institute of Science,
Rehovot, 76100, ISRAEL}
\centerline{and}
\centerline{Raymond and Beverly Sackler Faculty of Exact Sciences}
\centerline{School of Physics and Astronomy, Tel Aviv University,
Tel Aviv, Israel}
\bigskip
\medskip
\centerline{ABSTRACT}
\bigskip
\tenpoint
\baselineskip=12pt

{\noindent
      CP violation can be observed in B decays when a given process
depends upon interference between two weak amplitudes which have
different $CP$-violating phases. Since most weak decay diagrams have
quark lines where each has a definite flavor label, neutral mesons which
are flavor mixtures are particularly interesting. Different diagrams can
contribute to the different flavor components of the meson, and the wave
function itself provides interference.

\twelvepoint
\noindent
1.\  {\caps A Flavor Topology Classification}

\noindent
The total amplitude for
the decay of a $B$ meson consisting of a $b$ quark and an
antiquark denoted by $\bar q$ can be expressed as
as the sum of three independent amplitudes with different flavor
topologies\REF{\LipTop}{Harry J. Lipkin Phys.
Rev. Lett. 46 1307 (1981) }$[{\LipTop}]$:
$$
\vbox{\eqalignno{
B(b \bar q) & \rightarrow U_b W^- \bar q   \rightarrow U_b +
\bar U + D + \bar q                               &(QQ1)  \cr
B(b \bar q) & \rightarrow U_b W^- \bar q   \rightarrow U_b +
\bar U + f + \bar f                             &(QQ2a)  \cr
B(b \bar q) & \rightarrow U_b W^- \bar q   \rightarrow D +
\bar U + f + \bar f                             &(QQ2b)  \cr
B(b \bar q) & \rightarrow U_b W^- \bar q   \rightarrow
D + \bar q + f + \bar f                  &(QQ3)
\cr}}   $$
where $U$ denotes a quark of charge (+2/3),
$D$ a quark of charge (-1/3) and
$U_b$ denotes the $U$ quark produced in the initial $b \rightarrow  U$
transition with the emission of the $W^-$.
\noindent
The spectator tree diagram (QQ1) has all three flavors produced by the $
b$ decay present in the final state together with the spectator
antiquark $\bar q$.
The spectator annihilation diagrams (QQ2a) and (QQ2b) have the spectator
antiquark $\bar q$ annihilated either by the
$U_b$ or the $D$ quark produced in the tree diagram. These can arise
from a weak annihilation diagram, a weak $W$-exchange
diagram or a tree diagram followed by final state
interactions. Note
that (QQ2a) is allowed only if the spectator $\bar q$ is a $\bar c$ or
$\bar u$ and that (QQ2b) is allowed only if the spectator $\bar q$

\noindent
is a $\bar s$ or $\bar d$.
The penguin diagram (QQ3) contains a loop in which
the $U_b$ and $U$ annihilate and appear as a
single line in the weak penguin diagram. The same topology can arise
in a tree diagram followed by final state interactions.

\noindent
The particular products  of $CKM$ matrix elements in the standard
model that contribute to a given decay are completely determined by this
topological classification. The $CKM$ product arising in any
decay described by diagrams (QQ2) and (QQ3) is the same whether the
topology results directly from the weak diagram or from a tree diagram
followed by a final state interaction, with the exception of the flavor
label of the quark that is created and annihilated in the penguin and is
not directly observed. The topology alone
cannot determine the detailed dynamics, but can determine the particular
products of CKM matrix elements contributing to a particular decay.
The following flavor properties result from this classification:

\noindent
1. In the spectator tree diagram (QQ1)
the flavors of the four outgoing quark lines are determined
uniquely and completely by the weak vertex.

\noindent
2. In both the spectator annihilation (QQ2) and penguin (QQ3) diagrams a
flavor-symmetric
$f \bar f$
pair created by gluons appears in the final state.
These diagrams
come in triplets in which the additional pair is $u \bar u$,
$d \bar d$ and $s \bar s$
with amplitudes having
equal magnitudes and a positive relative phase in the SU(3) flavor-symmetry
limit. The $s \bar s$ amplitude is reduced by SU(3)
symmetry breaking due to the quark mass differences.
The flavor quantum numbers of
the final state are completely determined by the remaining
single $q \bar q$ pair with well defined non-exotic
flavor quantum numbers conserved in
all strong final state interactions.

\noindent
3. Final states having exotic flavor quantum numbers; i.e. containing
no $q \bar q$ pair of the same flavor, have only spectator tree
contributions.

\noindent
4. For final states containing no quark with the flavor quantum numbers of the
incoming spectator quark, only the spectator annihilation diagram can
contribute.

\noindent
5. For final states where both the spectator tree and spectator
annihilation diagrams contribute, $\bar f$, $\bar q$ and either $U_b$ or
$D$ must have the same flavor.
The tree diagram can therefore be turned into a spectator annihilation
diagram without changing the weak vertex by closing the lines of the
outgoing pair to make a loop annihilating the spectator quark and
creating an additional pair with gluons.
Thus both diagrams depend upon the same product of weak CKM matrix
elements in the standard model.

\noindent
6. A flavor-mixed neutral meson like $\eta$, $\eta`$, $\pi^o$, $\rho^o$ or
$\omega$ is produced only in final states containing a $q \bar q$ pair of
the same flavor. The spectator tree diagram generally contains only one
such $q \bar q$ pair and in that case produces the three neutral
pseudoscalars or vectors with a ratio determined completely by the amplitude of
the pair of that flavor in the wave function and corrections for phase space.
Thus any deviation from such a ratio; e.g. a larger $\eta'$ production than
$\eta$ production or unequal $\rho^o$ and $\omega$ production indicates
interference between the spectator tree diagram and one of the other two.

\noindent
7. The OZI rule forbids the creation of a flavor-mixed neutral meson like
$\eta$, $\eta`$, $\pi^o$, $\rho^o$ or $\omega$ from the $q \bar q$
pair of the same flavor produced by gluons in a quark loop in the spectator
annihilation or penguin diagram.

\noindent
8. If the final state contains an $f \bar f$ pair as in spectator
annihilation or penguin transitions, together with
a single light-quark $\bar q q$ pair
which is not flavor neutral and any number of additional gluons the final
state is a flavor-SU(3) octet in the SU(3) symmetry limit. When charge
conjugation quantum numbers forbid the SU(3) octet-octet-singlet
coupling; e.g. in the spin-zero pseudoscalar-vector final state, the OZI
rule is already predicted by SU(3) with no further assumptions.

\noindent
9. The only way to observe a CP-violating charge-asymmetric
quasi-two-body decay in
the standard model is by interference between two amplitudes depending upon
different CKM matrix elements. This requires interference involving the
penguin diagram\REF{\PBPENG}{Harry J. Lipkin,
Yosef Nir, Helen R. Quinn and Arthur E. Snyder,
Physical Review D44, 1454 (1991) 1454}$[{\PBPENG}]$.

\noindent
There is great interest in finding penguin contributions and CP-violating
interference effects, but as yet no firm experimental evidence for
penguins.
The use of final states containing $\eta$ and $\eta'$ has been
discussed\REF{\PKETA}{Harry J. Lipkin, Phys. Lett. B254
247, (1991) and B283 421, (1992)}$[{\PKETA}]$.
We concentrate here on newer predictions involving $\rho$ and
$\omega$, $D^o$ and $\bar D^o$.

\noindent
The $\rho^o$ and $\omega$ are equal mixtures with opposite relative phase
of the vector quarkonium states $u \bar u$ and $d \bar d$,
denoted respectively by $V_{u} $ and $V_{d}$. Equal $\rho$ and
$\omega$ production is predicted\REF{\ALS}{G. Alexander, H.J. Lipkin and
F. Scheck, Phys. Rev. Lett. 17, 412 (1966).}$[{\ALS}]$
for any quark diagram leading to a single flavor
state, either $V_{u} $ or $V_{d}$, together with
the interference effect\REF{\GLASHOW}{S. L. Glashow, Phys. Rev. Lett. 7,
469 (1961) }$[{\GLASHOW}]$ confirmed experimentally\REF{\GOLDHABER}{G.
Goldhaber, in Experimental Meson Spectroscopy, Edited
by Charles Baltay and Arthur H. Rosenfeld, Columbia University Press,
New York (1970)}$[{\GOLDHABER}]$.

\medskip

\noindent{ 2.\  {\caps
$B^o \rightarrow K^o \rho^o$ and $B^o \rightarrow K^o \omega$ Decays
}}

\noindent
In these
decays
the Cabibbo-suppressed color-suppressed spectator tree diagram
produces $V_u$, while
the penguin and all other diagrams via an intermediate
$\bar q q$ pair have the flavor quantum numbers $\bar s d $ and produce
$V_d$ in the transitions allowed by flavor SU(3) or OZI,
$$ \vbox{\eqalignno{
 B^o(\bar b d) & \rightarrow (\bar u u \bar s) d  \rightarrow K^o V_u
                     &(ZZ4a) \cr
  B^o(\bar b d) & \rightarrow \bar s d   \rightarrow  K^o V_d &(ZZ4b)
\cr}}   $$
$$ \vbox{\eqalignno{
 {{\tilde \Gamma (B^o \rightarrow K^o \omega)}\over{
\tilde \Gamma (B^o \rightarrow  K^o \rho^o )}}= &
\left| {{T + P}\over { T-P}}\right| ^2   &(ZZ5a) \cr
 {{\tilde \Gamma (\bar B^o \rightarrow \bar K^o \omega)}\over{
\tilde \Gamma (\bar B^o \rightarrow \bar K^o \rho^o )}}= &
\left| {{T + \bar P}\over { T-\bar P}}\right| ^2
 &(ZZ5b)
\cr}}   $$
where $\tilde \Gamma $ denotes the reduced partial width for the
particular decay mode with the dependence of the phase space factor
on individual final states removed,
T, and P denote the contributions to the decay amplitudes (ZZ5a)
from the tree and penguin diagrams respectively,
$\bar P$ denotes the penguin contribution to the charge conjugate
amplitudes (ZZ5b) and we use the Nir - Quinn phase
convention\REF{\NIRQUINN} {Yosef Nir and Helen Quinn Phys. Rev. Lett.
67, 541 (1991) }$[{\NIRQUINN}]$
in which the CP-violating phase appears only in the penguin diagram.

\noindent
This offers the possibility of detecting the penguin contribution and
also measuring the relative phase of penguin and tree contributions, as
well as detecting $CP$ violation in a difference between the
charge-conjugate $\rho/\omega$ ratios (ZZ5a) and (ZZ5b).
   $$ {{\tilde \Gamma (B^o \rightarrow K^o \omega)}\over{
\tilde \Gamma (B^o \rightarrow  K^o \rho^o )}}-
 {{\tilde \Gamma (\bar B^o \rightarrow \bar K^o \omega)}\over{
\tilde \Gamma (\bar B^o \rightarrow \bar K^o \rho^o )}}
\approx 4 Re\left[{{P - \bar P}\over{T}}\right]
 \eqno(ZZ6) $$

\noindent
Note, however, that tree diagrams followed by final state interactions
exist with the penguin topology and produce the $\rho$ and $\omega$
via an intermediate $\bar s d$ pair which decays into $K^o V_d$; e.g.
$$ \vbox{\eqalignno{
B^o \rightarrow & D^{*+} + \bar D_s \rightarrow
  K^o V_d &(ZZ7a)\cr
B^o \rightarrow  & K^+ + \rho^- \rightarrow
  K^o V_d &(ZZ7b)
\cr}}   $$
The diagram (ZZ7a) depends upon the same CKM matrix elements
$V_{bc}$ and $V_{cs}$  as the penguin and will have the same weak phase
in the standard model. Therefore the contribution of this diagram to the
amplitude $P$ in eqs. (ZZ5) will not
interfere with using these relations
to obtain CKM matrix elements. However
the diagram (ZZ7b) depends upon the same CKM matrix elements
$V_{bu}$ and $V_{us}$  as the tree diagram and will have
the same weak phase as the amplitude $T$ in eqs. (ZZ5), even though it
contributes to the $P$ amplitude in eqs. (ZZ5). This contribution thus
does interfere with using the relations (ZZ5) directly
to obtain CKM matrix elements, but does not affect the $CP$ violation
relation (ZZ6).

\noindent
Further information can be obtained by looking for the $\rho-\omega$
interference observed in strong reactions$[{\GOLDHABER}]$
in detailed analysis of the
$\pi^+ \pi^-$ spectrum over the mass range of the $\rho$ resonance.
The isospin violating $\omega \rightarrow \pi^+ \pi^-$ has a branching
ratio of only 2.2\%. But the width of the $\omega$ is 8.4 MeV while that
of the $\rho$ is 149 MeV\REF{\PDG}{Particle
Data Group, Phys. Lett. B239, 1 (1990).}
$[{\PDG}]$. Thus
   $$ {{\tilde \Gamma \{\omega
   \rightarrow (\pi^+ \pi^-)_\omega\}}\over{
\tilde \Gamma \{\rho^o \rightarrow (\pi^+ \pi^-)_\omega\}}} =
0.022 \cdot {149 \over 8.4} \approx 0.39
 \eqno(ZZ8) $$
where $(\pi^+ \pi^-)_\omega$ denotes the $\pi^+ \pi^-$ decay mode at the
$\omega$ peak. Thus
   $$ {{\tilde \Gamma \{B^o \rightarrow K^o \omega
\rightarrow K^o (\pi^+ \pi^-)_\omega\}
   }\over{
   \tilde \Gamma \{B^o \rightarrow K^o \rho^o
\rightarrow K^o (\pi^+ \pi^-)_\omega\}}}
= 0.39 \cdot
\left| {{ T + P}\over {  T- P}}\right| ^2
\eqno(ZZ9a) $$
   $$ {{\tilde \Gamma \{\bar B^o \rightarrow \bar K^o \omega
\rightarrow \bar K^o (\pi^+ \pi^-)_\omega\}
   }\over{
   \tilde \Gamma \{\bar B^o \rightarrow \bar K^o \rho^o
\rightarrow \bar K^o (\pi^+ \pi^-)_\omega\}}}
= 0.39 \cdot
\left| {{ T + \bar P}\over {  T- \bar P}}\right| ^2
\eqno(ZZ9b) $$
If the two contributions are coherent, the total contribution is given by
   $$ {{\tilde \Gamma \{\bar B^o
\rightarrow \bar K^o (\pi^+ \pi^-)_\omega\}
   }\over{
   \tilde \Gamma \{\bar B^o \rightarrow \bar K^o \rho^o
\rightarrow \bar K^o (\pi^+ \pi^-)_\omega\}}}
= $$ $$ =
1 + 1.25 \cos (\alpha+\phi_{PT})\cdot \left| {{T + P}\over {T- P}}\right|
+ 0.39\cdot
\left| {{ T + P}\over {  T- P}}\right| ^2
\eqno(ZZ10a) $$
where $\alpha$ is the relative phase of the $\rho$ and $\omega$
contributions and
$\phi_{PT}$ is a relative phase defined by
$$  {{ T + P}\over {  T- P}} \equiv
\left| {{ T + P}\over {  T- P}}\right| e^{i\phi_{PT}}
\eqno(ZZ10b) $$
and similarly for the charge conjugate case (ZZ9b).

\noindent
Additional information will be obtained
if enough statistics are available for observing the detailed behavior
of the decay as a function of energy through the resonance. The phase
$\alpha$ will change rapidly in passing through the $\omega$ resonance
and the interference pattern can give information on the phases
$\phi_{PT}$ and $\phi_{\bar P T}$. Any difference between the two
indicates $CP$ violation.

\noindent
Additional interference effects arise when the final kaon is detected in
the $K_S$ decay mode and the initial $B$ meson undergoes $B^o-\bar B^o$
oscillations\REF{\Bigi}{I. I. Bigi and A. I Sanda, Nuc. Phys.
B281, 41 (1987)}
\REF{\PEPRspin}{Harry J. Lipkin, Physics Letters B219, 474 (1989)
and ``Physics at B-Factories (A Theoretical Talk) these proceedings.}
$[{\Bigi,\PEPRspin}]$
There are two independent CP-violating relative phases: (1) the
relative phase of the $P$ and $\bar P$ amplitudes which expresses the
relative weak phase of the penguin and tree contributions; (2) a
parameter $\theta$ $[{\PEPRspin}]$
which expresses the weak phase contribution to the $B^o - \bar B^o$ mixing
relative to the phase of the tree contribution to the decays which has been
used to define the relative phase of the $B^o$ and $ \bar B^o$ states.
In addition there is the rapidly varying strong phase $\alpha$ (ZZ10a).
Thus measuring these decays both as a function of time and of the
invariant mass of the $\pi^+ \pi^-$ system can give interesting
information on decay amplitudes and $CP$ violation.

\medskip

\noindent{ 3.\ {\caps Cascade Decays via $D^o$ and
$\bar D^o$}}

\noindent
CP-violating charge asymmetries can arise in $B$ decays
via different diagrams involving different weak CKM matrix elements
into charmed states containing
$D^o$ and $\bar D^o$\REF{\GRONAU}{M. Gronau and D. Wyler, Phys. Lett.
B265, 172 (1991);
M. Gronau and D. London, Phys. Lett. B253, 483 (1991).}
\REF{\DUNIETZ}{I. Dunietz, Phys. Lett. B270, 74 (1991).}$[{\GRONAU,
\DUNIETZ}]$
which then decay into the same final state,
$$ \vbox{\eqalignno{
   B^\pm \rightarrow  & K^\pm + D^o \rightarrow
K^\pm + K^+ + K^-                      &(ZZ11a) \cr
B^\pm \rightarrow  & K^\pm + \bar D^o \rightarrow
K^\pm + K^+ + K^-                      &(ZZ11b)
\cr}}$$
Eq
(ZZ11a) involves $V_{bc}$ and $V_{su}$; eq.
(ZZ11b) involves $V_{bu}$ and $V_{sc}$.
There are also decays via $ K^\pm D^*$,
$$ B^\pm \rightarrow  K^\pm + D^{*o}(\bar D^{*o})
\rightarrow  K^\pm + \pi^o + K^+ + K^-              \eqno(ZZ12a) $$
$$ B^\pm \rightarrow  K^\pm + D^{*o}(\bar D^{*o})
\rightarrow  K^\pm + \gamma + K^+ + K^-             \eqno(ZZ12b)
$$
Since the $\pi^o$ and $\gamma$ have opposite $C$ the relative
phase of the $D^o$ and $\bar D^o$ produced in the two decay modes
are opposite. The opposite charge asymmetry predicted for the two cases
can provide a useful experimental check for systematic errors.

\medskip

\noindent{ 4.\  {\caps Charm Decays to $\rho$, $\omega$,
$\phi$}}

\noindent
Analogous flavor topologies can be defined in charm decays,
with flavor labels having the reverse weak isospin. By analogy with (ZZ5)
$$ \vbox{\eqalignno{
{\tilde \Gamma (D^o \rightarrow \bar K^o \omega )} & = (1/2)|T+A|^2
 &(ZZ13a) \cr
 {\tilde \Gamma (D^o \rightarrow \bar K^o \rho)}& = (1/2)|T-A|^2
 &(ZZ13b) \cr
 { \tilde \Gamma (D^o \rightarrow \bar K^o \phi)} & = \xi^2|A|^2
 &(ZZ13c) \cr}}  $$
where
$T$ and $A$ denote the contributions
from the tree and annihilation diagrams respectively,
and $\xi \leq 1$ is an SU(3)-breaking suppression factor.
The penguin does not contribute and there is no $CP$ violation asymmetry.
Experimental data$[{\PDG}]$ show a large difference
between $\rho$ and $\omega$ branching ratios and
an appreciable $\phi$, implying that the two amplitudes
$T$ and $A$ are comparable.
Combining eqs. (ZZ13) and introducing experimental values$[{\PDG}]$
give a triangular inequality
 $$ (1.61\pm 1.21)\%
\leq {{ \tilde BR_\omega (D^o \rightarrow \bar K^o \phi)}\over {2 \xi^2}}
\leq (6.65\pm 2.29)\% \eqno(ZZ14)$$
where $\tilde BR_\omega$ denotes that the branching ratio is normalized
to the $K-\omega$ phase space and needs a phase space correction.
The experimental value $(1/2) BR (D^o \rightarrow \bar K^o \phi) =
(0.4 \pm 0.08)\%$ might indicate a violation of the lower bound or large SU(3)
breaking. A better measurement of the $K \omega$ branching ratio would be
useful and a measurement of the relative phase of the $\omega$ and $\rho$
amplitudes via $\rho \omega$ interference would enable
a direct test of the relations (ZZ13).

An analogous treatment and flavor-topology
classification has been given for quasi-twobody $D^+$ and $D_s$
decays$[{\PKETA}]$
where $T$ denotes the sum of contributions from all diagrams in which the
spectator $\bar d$ quark remains in the final state and $A$ denotes the
contributions from
diagrams in which the spectator quark is annihilated and an additional
$q \bar q$ pair with all possible flavors is created by gluons.
\refout
\end